\documentclass[aps,preprint,eqsecnum,tightenlines]{revtex4}

\begin{document}

\title{Self-consistent system of equations for a kinetic description of the
low-pressure discharges accounting for the nonlocal and collisionless
electron dynamics}
\author{Igor D. Kaganovich }
\affiliation{Plasma Physics Laboratory, Princeton University, Princeton, NJ 08543}
\date{\today }

\begin{abstract}
For low-pressure discharges, when the electron mean free path is larger or
comparable with the discharge length, the electron dynamics is essentially
nonlocal. Moreover, the electron energy distribution function (EEDF)
deviates considerably from a Maxwellian. Therefore, an accurate kinetic
description of the low-pressure discharges requires knowledge of the
nonlocal conductivity operator and calculation of the nonMaxwellian EEDF.
The previous treatments made use of simplifying assumptions: a uniform
density profile and a Maxwellian EEDF. In the present study a
self-consistent system of equations for the kinetic description of nonlocal,
nonuniform, nearly collisionless plasmas of low-pressure discharges is
derived. It consists of the nonlocal conductivity operator and the averaged
kinetic equation for calculation of the nonMaxwellian EEDF. The importance
of accounting for the nonuniform plasma density profile on both the current
density profile and the EEDF is demonstrated.
\end{abstract}

\maketitle

\textbf{List of variables:}

$T_{b}$ is the bounce period,

$T$ is half of the bounce period, $T(\varepsilon
_{x})=T_{b}/2=\int_{x_{-}}^{x_{+}}dx/\left\vert v_{x}(x,\varepsilon
_{x})\right\vert $,

$\Omega _{b}$ is the bounce frequency, $\Omega _{b}=2\pi /T_{b}$ ,

$L$ is the gap width,

$R$ is half of the gap width, which is used to model a cylinder geometry, $%
R=L/2$,

$w$ is the electron kinetic energy, $%
w=w_{x}+w_{y}+w_{z}=m(v_{x}^{2}+v_{y}^{2}+v_{z}^{2})/2$,

$\varphi (x)$ is the electron potential energy, $\varphi (x)=-e\phi (x)$, and

$\phi $ is the electrostatic potential,

$\varepsilon $ is the total electron energy, $\varepsilon =w+\varphi $,

$\omega $ is the frequency of the rf electric field,

$\nu $ is the electron elastic collision frequency,

$\lambda $ is the electron elastic mean free path,

$\nu _{k}^{\ast }$ is the electron inelastic collision frequency for the
process number $k$,

$x_{-}(\varepsilon )$, $x_{+}(\varepsilon )$ are the left and right turning
points [$\varepsilon =\varphi (x_{\pm })$],

$\tau $ is the time of flight from the left turning point $x_{-}(\varepsilon
_{x})$ to $x$: $\tau (x,\varepsilon _{x})=\int_{x_{-}}^{x}dx/\left\vert
v_{x}(x,\varepsilon _{x})\right\vert $,

$\theta $ is the variable angle for bounce motion, defined as $\theta
(x)=\pi sgn(v_{x})\tau (x),/T(\varepsilon _{x})$

$\Phi (x,\varepsilon _{x})$ is the generalized phase of the rf electric
field, $\Phi (x,\varepsilon _{x})=\int_{x_{-}}^{x}(-i\omega +\nu
)dx/\left\vert v_{x}(x,\varepsilon _{x})\right\vert $. If $\omega >>\nu $, $%
\Phi \simeq \omega \tau $,

$\delta $ is the width of the skin layer,

$f$ is the electron velocity distribution function (EVDF), $%
f=f_{0}(\varepsilon )+f_{1}$,where $f_{0}(\varepsilon )$ is the main part of
the electron velocity distribution function averaged over velocity
directions and over available space for electrons with a given total energy $%
\varepsilon $, which is referred in the following as the electron energy
probability function (EEPF). Notwithstanding the fact that $f$ is defined in
the velocity space, we shall look for $f(\varepsilon )$ as a function of the
energy. The EEPF is normalized as $n=\int fd^{3}v=4\pi \sqrt{2}%
/m^{3/2}\int_{\varphi (x)}^{\infty }f_{0}(\varepsilon )\sqrt{\varepsilon
-\varphi (x)}d\varepsilon $, where $n$ is the electron density, and the
factor $4\pi \sqrt{2}/m^{3/2}$ is included in the definition of $f_{0}$ for
convenience. $f_{1}$ is the rapidly varying, anisotropic part of the EVDF,

$E_{sc}(x)$ is the space charge stationary electric field,

$E_{y}(x,t)$ is the rf nonstationary electric field,

$St(f)$ is the collision integral,

$V^{rf}$ is the oscillatory electron velocity driven by rf electric field.

\section{Introduction}

Low pressure radio-frequency discharges are extensively utilized for plasma
processing and lighting \cite{Liebermann book}. Simulation of discharge
properties is a common tool for optimization of the plasma density profiles
and ion and electron fluxes. Recent plasma technology tends to decrease the
gas pressures down to the millitorr range. For these low pressures it is
easier to maintain uniform plasmas with well controlled parameters. Due to
the large value of the electron mean free path ($\lambda $) the electron
current is determined not by the local rf electric field (Ohm's law), but
rather is a function of the whole profile of the rf electric field on
distances of order $\lambda $ (anomalous skin effect). Therefore, a rather
complicated nonlocal conductivity operator has to be determined for the
calculation of the rf electric field penetration into the plasma. Moreover,
the electron energy distribution function (EEDF)\ is typically nonMaxwellian
in these discharges \cite{Godyak new exp}. Hence, for accurate calculation
of the discharge characteristics at low pressures, the EEDF needs to be
computed self-consistently. Self-consistency is an important and difficult
issue for the kinetic simulations of a plasma. The EEDF, nonlocal
conductivity and plasma density profiles are all nonlinear and nonlocally
coupled. That is why, the self-consistency aspect of the model is the main
concern of this study. The so-called "nonlocal" approach relies on the
direct semi-analytic solution of the Boltzmann equation in the limiting
regime where the electron relaxation length is much large than the discharge
gap, but the electron mean free path is small compared with the discharge
dimension \cite{Bernstein}, \cite{Tsendin 1972}. The nonlocal approach has
been successfully applied to the self-consistent kinetic modelling of
various low-pressure discharges, where the electron mean free path is small:
the capacitively coupled plasmas \cite{My CCP}; \cite{My CCP gamma}, the
inductively coupled plasmas \cite{Kolobov ICP}, \cite{Kolobov 2ICP}, \cite%
{Uwe's ICP}, \cite{Muemkin}; the dc discharges \cite{Kolobov dc}, \cite%
{Benke}; the afterglow \cite{Robert}, and the surface-wave discharges \cite%
{Shlueter}. The additional references can be found in reviews: \cite{Tsendin
review}, \cite{Kolobov and Godyak review}, and \cite{Uwe review}.

If gas pressure lowered even farther, the electron mean free path becomes
comparable or even larger than the discharge dimension and numerous
collisionless phenomena dominate the discharge characteristics \cite{Lieber
& Godyak review}. In the present paper the nonlocal approach is generalized
for the low-pressure discharges to account for the collisionless heating and
transit-time (electron inertia) effects in the discharge description.

Present analysis considers only an inductively coupled plasma. But the
approach has been designed in the most generalized way, so that derivations
can be readily performed for other discharges. For example, in Ref.\cite{My
PRL 2002} the capacitive discharge; in Ref. \cite{My ECR} the
electron-cyclotron-resonance discharge and in Ref.\cite{Uwe's surface}
surface-wave discharge were considered with self-consistent account for
collisionless heating.

Most previously reported theoretical studies assume a uniform plasma, in a
semi-infinite \cite{Weibel} or a slab geometry \cite{Blevin theory}. In this
case the analytical treatment considerably simplifies, because electron
trajectories are straight. In the semi-infinite case, electrons traverse the
region of the rf electric field (skin layer) and are reflected back into the
plasma at the discharge walls. An acquired velocity kick then dissipates in
the plasma on distances of order the electron mean free path. If the plasma
dimension is small or comparable with $\lambda $, the subsequent kicks are
correlated. The resonance between the wave frequency and the bounce
frequency of the electron motion between walls may result in modification of
the nonlocal conductivity \cite{Blevin}, \cite{Shaing} and may yield an
enhanced electron heating \cite{me APL}. The anomalous skin effect has been
studied experimentally in cylindrical \cite{Blevin} and planar discharges
\cite{Godyak exp}. Additional references can be found in the reviews of
classical and recent works on the anomalous skin effect in gas discharge
plasmas \cite{Kolobov review}, \cite{F.F. Chen}. The theoretical studies in
cylindrical geometry are much more cumbersome, and has been done for uniform
plasma in Refs. \cite{Meierovich 4}, \cite{Yoon}, \cite{Kolobov PRE 55} and
for a parabolic potential well in Ref.\cite{Storer}. Qualitative results in
the cylindrical geometry are similar to the results in the plane geometry,
therefore, in the present study only one-dimensional slab geometry is
considered.

For the case of a bounded uniform plasma, the electrostatic potential well
is flat in the plasma and infinite at the wall (to simulate the existence of
sheaths). In this square potential well, electrons are reflected back into
the plasma only at the discharge walls. In a realistic non-uniform plasma,
however, the position of the turning points will depend on the electron
total (kinetic plus potential) energy and the actual shape of the potential
well, i.e., low total energy electrons bounce back at locations within the
plasma and may not reach regions of high electric field at all. As a result
the current density profiles in a nonuniform plasma may considerably differ
from the profiles in a uniform plasma. The theory of the anomalous skin
effect for an arbitrary profile of the electrostatic potential and a
Maxwellian EVDF was developed by Meierovich et al. in Refs.\cite{Meierovich}%
, \cite{Meierovich 2}, and \cite{Meierovich 3} for the slab geometry.
Although rigorous analytical results of non-uniform plasmas have been
reported, the detailed self-consistent simulations related to such plasmas
and comparison with experimental data are lacking. Such simulations were
completed recently and presented in our separate publications \cite{badri
and me} and \cite{badri and me 2} and will be additionally reported
elsewhere.

The kinetic description of the anomalous skin effect is based on a well
known mechanism of collisionless power dissipation -the Landau damping \cite%
{Landau}. In the infinite plasma, the resonance particles moving with a
velocity ($\mathbf{v}$) close to the wave phase velocity, so that $\omega =%
\mathbf{v\cdot k}$, intensively interact with wave fields. Therefore, the
collisionless electron heating (and the real part of the surface impedance)
depends on the magnitude of a Fourier harmonic of the electric field [$E(k)$%
] and the number of the resonant particles [$f(v_{x}=\omega /k),\mathbf{x=k/}%
k$]. That is why, the momentum acquired in the skin layer of width $\delta $
is maximal if the projection of velocity perpendicular to the plasma
boundary ($x$ - axis direction) is of order $\omega /\delta $. If the
interaction with the skin layer are repeated in a resonance manner the
momentum changes are mounted up. Therefore, the main contribution to the
electron heating and the resistive part of the surface impedance comes from
these resonant electrons. In a bounded plasma, the resonance condition
requires the bounce period ($T_{b}$) be equal to one or several rf electric
field periods: $T_{b}=2\pi n/\omega $, where $n$ is an integer number. The
maximum interaction occurs for $n=1$ (see below). For a slab of width $L$, $%
T_{b}=2L/v_{x}$.\ The maximum electron heating occurs if both aforementioned
conditions are satisfied simultaneously, which gives $\omega /\delta =v_{x}$
and $2L/v_{x}=2\pi /\omega $ or $L=\delta \pi $ \cite{Shaing}. \emph{Hence,
the optimum conditions for the power transfer to the plasma corresponds to
the plasma of size comparable with the 3 times of the skin depth.} Because
the bounce frequency depends on the electrostatic potential, accounting for
the plasma nonuniformity is important for a correct calculation of the
efficient power coupling.

As discussed before, the collisionless heating is determined by the number
of resonant particles, and, hence, is dependent on the EEDF. The EEDF, in
its turn, is controlled by the collisionless heating. The only particles,
which are in resonance with a wave, are heated by the collisionless heating.
It means that in the regime of the collisionless dissipation, the form of
the electron energy distribution function is sensitive to the wave spectrum.
Therefore, the plateau in the EEDF can be formed in the regions of intensive
collisionless heating, if the wave phase velocities are confined in some
interval \cite{ved67}. The evidences of a plateau formation for the
capacitive discharge plasma were obtained in Ref. \cite{Ulrich and me}. The
cold electrons, which are trapped in the discharge center, do not reach
periphery plasma regions where an intensive rf elective field is located,
and as a result, these electrons are not heated by the rf electric field.
The coupling between the EEDF shape and collisionless heating may result in
new nonlinear phenomena: an explosive generation of the cold electrons \cite%
{cold electron formation}. The experimental evidences of the collisionless
heating on the EEDF were obtained in Ref. \cite{Godyak exp}, \cite{Godyak
EDF}, \cite{Chin WOOK}and \cite{Chin Wook EDF influence of bounce}.

In the linear approximation the collisionless dissipation does not depend
explicitly on the collision frequency. However, as shown in Ref. \cite{Me
PRL 1998}, if the electron elastic collision frequency is too small, heating
can actually decrease due to nonlinear effects akin of the nonlinear Landau
damping.

The present article presents a self-consistent system of equations
describing the non-local electron kinetics in a 1-D slab (bounded)
non-uniform plasma. The system consists of a nonlocal conductivity operator,
and an averaged over fast electron motions kinetic equation for the EEDF.
Transit time (non-local) effects on the current density profile and
collisionless heating are of particular interest. Rigorous derivations for
the nonlocal conductivity operator have been performed. The analytic results
of Ref.\cite{Meierovich} for the Maxwellian EEDF were generalized for the
nonMaxwellian EEDF. The spectral method was developed to find the rf
electric field profile. A quasilinear approach was used for calculating the
collisionless heating. The quasilinear theory developed in Ref.\cite{Aliev
and me} was generalized for an arbitrary value of the collision frequency.
As a result, the simulations can be done in a wide range of the background
gas pressures ranging from the collisional case ($\lambda <<\delta $) to the
fully collisionless case ($\lambda >L$). Self-consistency of the nonlocal
conductivity operator and the energy diffusion coefficient has been
verified: both yield the same expression for the power deposition. The
robust time-averaging procedure was designed for the kinetic equation in a
most general way. As a result, the procedure can be readily repeated for
other discharges, see for example \cite{My PRL 2002} and \cite{My ECR}.

\section{Calculation of anisotropic part of the electron velocity
distribution function $f_{1}$}

In low-pressure discharges, where the energy relaxation length is large
compared with the plasma width, the main of the electron velocity
distribution function (EVDF) is a function of the total energy only \cite%
{Tsendin review}, \cite{Kolobov and Godyak review} and \cite{Uwe review}.
Therefore, we look for $f=f_{0}+f_{1}$, where $f_{0}(\varepsilon )$ is a
function of the total energy $\varepsilon $, $\varepsilon =w+\varphi (x)$, $%
w=m(v_{x}^{2}+v_{y}^{2}+v_{z}^{2})/2\ $is\ the kinetic\ energy, $\varphi
=-e\phi $ is the electron electrostatic potential energy, and $\phi $ is the
electrostatic potential. $f_{1}$ does not contribute to the electron density
(the integral $f_{1}$ over the velocity space is equal to zero $\int
f_{1}d^{3}v=0$), but $f_{1}$ contributes to the electron current (the
integral $f_{0}$ over the velocity space weighted with the electron velocity
is equal to zero $\int f_{0}\mathbf{v}d^{3}v=0$). Typically the mean
electron flow velocity ($V^{rf}=\int \mathbf{v}f_{1}d^{3}v/\int f_{0}d^{3}v$%
) is small compared with the thermal velocity ($V_{T}\equiv \sqrt{2T/m}$).
Therefore, the isotropic part of the EVDF is larger than the anisotropic
part $f_{1}\sim (V^{rf}/V_{T})f_{0}<<f_{0}$ \cite{Tsendin review}, \cite%
{Kolobov and Godyak review}, and \cite{Uwe review}.

Vlasov's equation reads:

\begin{equation}
\frac{\partial f_{1}}{\partial t}+v_{x}\frac{\partial f_{1}}{\partial x}+%
\frac{eE_{sc}(x)}{m}\frac{\partial f_{1}}{\partial v_{x}}+\frac{eE_{y}(x,t)}{%
m}\frac{\partial (f_{0}+f_{1})}{\partial v_{y}}=St(f_{1}+f_{0}),
\label{Vlasov eq. 0}
\end{equation}%
where $E_{sc}(x)$ is the space-charge stationary electric field, and $%
E_{y}(x,t)$ is the rf nonstationary electric field, $St(f)$ is the collision
integral. In the Eq.(\ref{Vlasov eq. 0}), we used the fact that
\begin{equation}
v_{x}\frac{\partial f_{0}(\varepsilon )}{\partial x}+\frac{eE_{sc}(x)}{m}%
\frac{\partial f_{0}(\varepsilon )}{\partial v_{x}}=v_{x}\frac{\partial
f_{0}(\varepsilon )}{\partial x}|_{\varepsilon _{x}}=0,
\end{equation}%
because $\varepsilon _{x}$ is constant along a trajectory. After applying
the standard quasilinear theory, Eq.(\ref{Vlasov eq. 0}) splits into two
equations \cite{Aliev and me}: a linear equation for $f_{1}$

\begin{equation}
\frac{\partial f_{1}}{\partial t}+v_{x}\frac{\partial f_{1}}{\partial x}+%
\frac{eE_{sc}(x)}{m}\frac{\partial f_{1}}{\partial v_{x}}+\frac{eE_{y}(x,t)}{%
m}\frac{\partial f_{0}}{\partial v_{y}}=St(f_{1}),  \label{Eq. for f1}
\end{equation}%
and a quasilinear equation for $f_{0}$

\begin{equation}
\overline{\frac{eE_{y}(x,t)}{m}\frac{df_{1}}{dv_{y}}}=\overline{St(f_{0})},
\label{Eq. for f0}
\end{equation}%
where upper bar denotes space-time averaging over the phase space available
for the electron with the total energy $\varepsilon $ \cite{Tsendin 77 dc},
\cite{Me and Tsendin 1992 1},\cite{Me and Tsendin 1992 2}.

The rf electric field $E_{y}(x,t)=E_{y0}(x)\exp (-i\omega t)$ and the
anisotropic part of the EVDF $f_{1}=f_{10}\exp (-i\omega t)$ are harmonic
functions, where $\omega $ is the discharge frequency. In what follows the
subscript $0$ is omitted. Eq.(\ref{Eq. for f1}) becomes
\begin{equation}
-i\omega f_{1}+v_{x}\frac{\partial f_{1}}{\partial x}|_{\varepsilon
_{x}}+ev_{y}E_{y}(x)\frac{df_{0}}{d\varepsilon }=-\nu f_{1}.
\label{Vlasov 1}
\end{equation}%
In transformation from Eq.(\ref{Eq. for f1}) to Eq.(\ref{Vlasov 1}) the BGK
approximation was used $St(f_{1})=-\nu f_{1}$,\ where $\nu $ is the
transport collision frequency and introduced the new variable: the total
energy along $x$-axis $\varepsilon _{x}=mv_{x}^{2}/2+\varphi (x)$. There
have been a number of studies, which explored the effects of the exact
collision integral on collisionless phenomena \cite{WW Lee}, \cite{Furkal}.
These treatments use expansion in series of spherical functions in velocity
spaces. The exact calculation are important only if the collision frequency
is a strong function of the polloidal scattering angle. If the differential
cross section does not depend on the polloidal scattering angle, the BGK
approximation is correct exactly \cite{Furkal}. For partially ionized plasma
electron-neutral collisions are the most frequent scattering mechanism. At
typical electron energies in the low-pressure discharges 1-5 eV \cite{Godyak
new exp}, the differential cross section weakly depends on the polloidal
scattering angle, and, therefore, the BGK approximation has a good accuracy.

Equation (\ref{Vlasov 1}) can be solved by a number of different methods.
First, let us consider a direct solution. Alternative derivation using
Fourier series is performed in Appendix C. After some straightforward
algebra described in Appendix A, the symmetric part of the EVDF $%
f_{1s}\equiv 1/2(f_{1+}+f_{1-})$ is given by

\begin{equation}
f_{1s}(\mathbf{v,x)}=-mv_{y}V_{y}^{rf}(x,\varepsilon _{x})\frac{df_{0}}{%
d\varepsilon },  \label{f1s paper}
\end{equation}%
where $V_{y}^{rf}(x,\varepsilon _{x})=1/2(V_{y+}^{rf}+V_{y-}^{rf})$, $%
V_{y\pm }^{rf}$ are the oscillatory velocities of an electron with a given $%
\varepsilon _{x}$, $\pm $ signs denote $v_{x}>0$ and $v_{x}<0$,
respectively;
\begin{equation}
V_{y}^{rf}(x,\varepsilon _{x},v_{\perp })=\frac{e}{m\sinh \Phi _{+}}\left[
\begin{array}{c}
\cosh \Phi \int_{x}^{x_{+}}E_{y}(x^{\prime })\cosh (\Phi _{+}-\Phi ^{\prime
})d\tau ^{\prime }+ \\
\cosh (\Phi _{+}-\Phi )\int_{x_{-}}^{x}E_{y}(x^{\prime })\cosh \Phi ^{\prime
}d\tau ^{\prime }%
\end{array}%
\right] ,  \label{Vrf(x,e)}
\end{equation}%
\begin{equation}
\tau \equiv \int_{x_{-}}^{x}\frac{dx}{\left\vert v_{x}(x,\varepsilon
_{x})\right\vert },  \label{tau}
\end{equation}%
\begin{equation}
\Phi (x,\varepsilon _{x},v_{\perp })\equiv \int_{x_{-}}^{x}(-i\omega +\nu
)d\tau ,  \label{phase paper}
\end{equation}%
\begin{equation}
\Phi _{+}(\varepsilon _{x},v_{\perp })\equiv \Phi (x_{+},\varepsilon
_{x},v_{\perp }),  \label{phase full paper}
\end{equation}%
where $x_{-}(\varepsilon _{x}),x_{+}(\varepsilon _{x})$ are the left and
right turning points, respectively, for the electron with energy $%
\varepsilon _{x}$ [corresponding to zero velocity $v_{x}$ or $\varepsilon
_{x}=e\varphi (x_{-})$], $\tau $ is the time of flight from the left turning
point $x_{-}(\varepsilon _{x})$ to $x$, $v_{\perp }=\sqrt{v_{y}^{2}+v_{z}^{2}%
}$. The functions $V_{y}^{rf}$ and $\Phi $ depend on the electron speed via
the collision frequency $\nu (v)$.

In the local limit the electron mean free path is large $\lambda
>>\delta $ and phase $Im(\Phi )>>1$. Therefore, $\cosh \Phi
\approx \sinh \Phi
\approx 1/2\exp (\Phi )$. The main contribution in the both integrals in Eq.(%
\ref{f1s final}) are near the point $x^{\prime }=x$ and hence $d\Phi
=(-i\omega +\nu )d\tau $
\begin{equation}
V_{y}^{rf}\approx \frac{e}{m}\frac{E_{y}(x)}{-i\omega +\nu },
\label{Vrf local}
\end{equation}%
as it should be in the local limit.

\section{Calculation of nonlocal conductivity}

Knowing the EVDF $f_{1s}$, one can calculate the current density

\begin{equation}
j=\frac{em^{3/2}}{4\pi \sqrt{2}}\int f_{1s}v_{y}d^{3}\mathbf{v}.
\label{J definiton}
\end{equation}%
Substituting $f_{1s}$ from Eq.(\ref{f1s paper}) into Eq.(\ref{J definiton})
and making the transformation to the spherical coordinates in the velocity
space $dv_{x}dv_{y}dv_{z}=v^{2}dv\sin \vartheta d\vartheta d\psi $ in ($\cos
\vartheta =v_{x}/v;$ $\tan \psi =v_{y}/v_{z}$ ) Eq.(\ref{J definiton})
becomes
\begin{equation}
j(x)=-e\sqrt{2}m^{3/2}\int_{0}^{\infty }w\left\langle
v_{y}^{2}V_{y}^{rf}\right\rangle \frac{df_{0}(\varepsilon )}{d\varepsilon }%
dv,  \label{j(G)}
\end{equation}%
where the averaged over velocity direction factor $\left\langle
v_{y}^{2}V_{y}^{rf}\right\rangle $ is
\begin{equation}
\left\langle v_{y}^{2}V_{y}^{rf}\right\rangle \equiv \frac{v^{2}}{4\pi }%
\int_{0}^{\pi }\int_{0}^{2\pi }V_{y}^{rf}(x,\varepsilon _{x},v)[\sin
\vartheta ]^{3}[\cos \psi ]^{2}d\psi d\vartheta .  \label{G1}
\end{equation}%
Because $V_{y}^{rf}$does not depend on $\psi ,$ the integration over $\psi $%
-angle can be completed. Changing integral from $\vartheta $ to $v_{x}=v\cos
\vartheta $ gives
\begin{equation}
\left\langle v_{y}^{2}V_{y}^{rf}\right\rangle \equiv \frac{1}{4v}%
\int_{-v}^{v}V_{y}^{rf}(x,\varepsilon _{x},v_{\perp
})(v^{2}-v_{x}^{2})dv_{x},  \label{G2}
\end{equation}%
or
\begin{equation}
\left\langle v_{y}V_{y}^{rf}\right\rangle \equiv \frac{1}{2m\sqrt{w}}%
\int_{\varphi (x)}^{\varepsilon }V_{y}^{rf}(x,\varepsilon _{x},v_{\perp })%
\frac{\varepsilon -\varepsilon _{x}}{\sqrt{\varepsilon _{x}-\varphi (x)}}%
d\varepsilon _{x}.  \label{G3}
\end{equation}%
Substituting Eq.(\ref{G3}) into Eq.(\ref{j(G)}) and changing integration
from $v$ to $\varepsilon $ yields
\begin{equation}
j(x)=-\frac{e}{2}\int_{0}^{\infty }\left[ \int_{\varphi (x)}^{\varepsilon }%
\frac{\varepsilon -\varepsilon _{x}}{\sqrt{\varepsilon _{x}-\varphi (x)}}%
V_{y}^{rf}d\varepsilon _{x}\right] \frac{df_{0}(\varepsilon )}{d\varepsilon }%
d\varepsilon .  \label{j(G)2}
\end{equation}%
Further simplifications are possible if the collision frequency $\nu $ is
small ($\nu <<\omega $) or $\nu $ does not depend on electron velocity. In
this case $V_{y}^{rf}(x,\varepsilon _{x},v_{\perp })$ is the only function
of $(x,\varepsilon _{x})$. Integrating Eq.(\ref{j(G)2}) in parts yields
\begin{equation}
j(x)=\frac{e}{2}\int_{\varphi (x)}^{\infty }\left[ \int_{\varphi
(x)}^{\varepsilon }\frac{V_{y}^{rf}(x,\varepsilon _{x})}{\sqrt{\varepsilon
_{x}-\varphi (x)}}d\varepsilon _{x}\right] f_{0}(\varepsilon )d\varepsilon .
\label{j(G)1}
\end{equation}%
If $V_{y}^{rf}$ is a constant Eq.(\ref{j(G)1}) gives trivial result: $%
j=enV_{y}^{rf}$.

Introducing a new function $\Gamma (\varepsilon )$
\begin{equation}
\Gamma (\varepsilon )\equiv \int_{\varepsilon }^{\infty }f_{0}(\varepsilon
)d\varepsilon ,  \label{definition of Gamma}
\end{equation}%
and integrating Eq.(\ref{j(G)1}) in parts one more time gives%
\begin{equation}
j(x)=\frac{e}{2}\int_{\varphi (x)}^{\infty }\frac{V_{y}^{rf}(x,\varepsilon
)\Gamma (\varepsilon )}{\sqrt{\varepsilon -\varphi (x)}}d\varepsilon
\label{j(A)final}
\end{equation}%
For the Maxwellian EVDF $f_{0}$, Eq.(\ref{j(A)final}) is equivalent to
Lieberman's et \textit{al.} result \cite{Meierovich}.

Substituting Eq.(\ref{Vrf(x,e)}) into Eq.(\ref{j(A)final}) yields the
nonlocal conductivity operator%
\begin{equation}
j_{y}(x)=\int_{0}^{x}G(x,x^{\prime })E_{y}(x^{\prime })dx^{\prime
}+\int_{x}^{L}G(x^{\prime },x)E_{y}(x^{\prime })dx^{\prime }
\label{J=G(xx')}
\end{equation}

where

\begin{equation}
G(x,x^{\prime })=\frac{1}{2}\frac{e^{2}}{\sqrt{2m}}\int_{\max (\varphi
,\varphi ^{\prime })}^{\infty }\frac{\cosh \Phi \cosh (\Phi _{+}-\Phi
^{\prime })}{\sinh \Phi _{+}}\frac{\Gamma (\varepsilon )}{\sqrt{\varepsilon
-\varphi (x)}\sqrt{\varepsilon -\varphi (x^{\prime })}}d\varepsilon .
\label{G(x,x')}
\end{equation}%
Note that $G(x,x^{\prime })$ has a logarithmic singularity at $x=x^{\prime }$
\cite{Meierovich}, but because calculation of the electron current in Eq.(%
\ref{J=G(xx')}) requires additional integration, there is no singularity in
the current.

In the limit of large gap, where $\delta <\lambda <<L$, $Re(\Phi
)>>1$ and $\cosh \Phi \cosh (\Phi _{+}-\Phi ^{\prime })/\sinh \Phi
_{+}\rightarrow 1/2[\exp (\Phi -\Phi ^{\prime })+\exp (-\Phi -\Phi
^{\prime })]$. And the region of integration in
Eq.(\ref{J=G(xx')}) beyond the skin layer can be omitted. In the
local limit, where $\lambda <<\delta $ , Eq.(\ref{J=G(xx')}) gives
the standard local conductivity, see Eq.(\ref{Vrf local}).

\section{Calculation of the transverse rf electric field profile}

Maxwell's equations can be reduced to a single scalar equation for the
transverse electric field \cite{Kolobov review}

\begin{equation}
\frac{d^{2}E_{y}}{dx^{2}}+\frac{\omega ^{2}}{c^{2}}E_{y}=-\frac{4\pi i\omega
}{c^{2}}\left[ j(x)+I\delta (x)-\delta _{anti}I\delta (x-L)\right] ,
\label{Maxwell eqs}
\end{equation}%
where the electron current $j$ is given by Eq.(\ref{J=G(xx')}), $I$ is the
current in the coil , $\delta _{anti}=0$, if the there is no any coil with
the current located at $x=L,$ and $\delta _{anti,k}=1$, if there is a coil
with the current $-I$ at $x=L$. The 1D slab system of two currents flowing
in opposite directions describes very well a cylindrical configuration,
where a coil produces rf currents at both plasma boundaries $x=-R$ and $x=R,$
$R=L/2$ \cite{Blevin theory}, \cite{Blevin}. The Eq.(\ref{Maxwell eqs}) and
Eq.(\ref{J=G(xx')}) can be solved numerically using a finite difference
scheme. There is major difficulty in such approach. Straightforward
computing the complex Green's function in Eq.(\ref{G(x,x')}) is slow and
time consuming \cite{badri and me}. The better approach is to solve the
integro-differential Eq.(\ref{Maxwell eqs}) making use of a spectral method.
In 1D geometry the electric field can be represented as a sum of harmonic
functions. The method is described in the Appendix D.

\section{Averaging of kinetic equation for the main part of the EVDF}

Kinetic equation for $f_{0}$ averaged over the discharge period is

\begin{equation}
v_{x}\frac{\partial f_{0}}{\partial x}+\frac{e}{m}E_{sc}(x)\frac{df_{0}}{%
dv_{x}}+\frac{e}{2m} Re\left[ E_{y}^{\ast }(x)\frac{df_{1}}{dv_{y}}%
\right] =St(f_{0}),  \label{Kinetic equation for f_0 1}
\end{equation}%
\begin{equation}
St(f)=St_{el}^{\mathbf{v}}(f)+St_{el}^{\varepsilon
}(f)+St_{ee}(f)+St_{inel}(f),  \label{collision integral general}
\end{equation}%
\begin{equation}
St_{el}^{\mathbf{v}}(f)=\int (f^{\prime }-f)d\sigma ,
\label{elastic scattering}
\end{equation}%
\begin{equation}
St_{el}^{\varepsilon }(f)=\frac{\partial }{v\partial w}\left(
vV_{el}f\right) ,  \label{elastic energy}
\end{equation}%
\begin{equation}
St_{ee}(f)=\frac{\partial }{v\partial w}\left( vD_{ee}\frac{\partial }{%
\partial w}f\right) +\frac{\partial }{v\partial w}\left( vV_{ee}f\right) ,
\label{electron electron collision}
\end{equation}%
\begin{equation}
St_{inel}(f_{0})=\sum_{k}\left[ \frac{\sqrt{(w+\varepsilon _{k}^{\ast })}}{%
\sqrt{w}}\nu _{k}^{\ast }f_{0}(w+\varepsilon _{k}^{\ast })-\nu _{k}^{\ast
}f_{0}\right] ,  \label{inelastic collision integral}
\end{equation}%
where $w=mv^{2}/2$ is the kinetic energy, $St_{el}^{\mathbf{v}}(f)$ is the
part of the elastic scattering collision integral with differential cross
section $d\sigma $, which changes the electron momentum but does not alter
the kinetic energy, $St_{el}^{\varepsilon }(f)$ accounts for energy change
in elastic collisions, $St_{ee}(f_{0})$ is the electron-electron collision
integral, and $St_{inel}(f_{0})$ is the sum over all inelastic collisions
with the electron energy loss $\varepsilon _{k}^{\ast }$ and inelastic
collision frequency $\nu _{k}^{\ast }$ (see details for ionization and wall
losses in \cite{My CCP}, \cite{My ECR}). Here, the coefficients $%
D_{ee},V_{ee}$, $V_{el}$ are given by \cite{Gurevich}, \cite{Me and Tsendin
1992 1},%
\begin{equation}
V_{el}=\frac{2m}{M}w\nu ,  \label{vel}
\end{equation}%
\begin{equation}
V_{ee}=2w\nu _{ee}\frac{\int_{0}^{w}dw\sqrt{w}f}{n},  \label{Vee}
\end{equation}%
\begin{equation}
D_{ee}=\frac{4}{3}w\nu _{ee}\frac{\int_{0}^{w}dww^{3/2}f+w^{3/2}\int_{w}^{%
\infty }dwf}{n},  \label{Dee}
\end{equation}%
\begin{equation}
\nu _{ee}=\frac{4\pi \Lambda _{ee}n}{m^{2}v^{3}},  \label{nu ee}
\end{equation}%
where $\nu _{ee}$ is the Coulomb collision frequency and $\Lambda _{ee}$ is
the Coulomb logarithm. Note that at large electron energies $V_{ee}=2w\nu
_{ee}$ and $D_{ee}=2wT_{e}\nu _{ee}$ , where $T_{e}=2/3%
\int_{0}^{w}dww^{3/2}f/n$.

If the electron mean free path is large compared with gap ($\lambda >>L$),
the first two terms on the left hand side are dominant. Therefore, $f_{0}$
is approximately a function of the $\varepsilon _{x}$ only, not a function
of both variables $x,v_{x}$ separately. Similarly, $St_{el}^{\mathbf{v}}(f)$
is the largest term from the remaining terms in the equation. Therefore, $%
f_{0}$ is approximately isotropic, and is a function of $\varepsilon $ only,
so that $\ St(f_{0})=0$ \cite{Me and Tsendin 1992 2}. This assumption was
verified by comparison with particle-in-cell simulations in Ref.\cite{My CCP}
for a capacitive coupled plasma, in Ref.\cite{Kolobov 2ICP}, \cite{Kolobov
PRE 55} for a inductive coupled plasma, and in Ref.\cite{My ECR} for a ECR
discharge and in Ref.\cite{Chin Wook EDF total e check} experimentally.

To find $f_{0},$ it is necessary to average over fast electron bouncing and
over all velocity angles. First, let us average over fast electron bouncing.
In order to do so, we integrate all terms of Eq.(\ref{Kinetic equation for
f_0 1}) over the full period of electron bouncing
\[
\oint dtTerm(x,v_{x})\equiv \int_{x_{-}}^{x_{+}}\frac{dx}{v_{x}}%
Term(x,v_{x}>0)+\int_{x_{-}}^{x_{+}}\frac{dx}{|v_{x}|}Term(x,v_{x}<0),
\]%
where $Term(x,v_{x})$ is a term in Eq.(\ref{Kinetic equation for f_0 1}).
Because the first two terms represent the full time derivative $df/dt$ along
trajectory, they disappear after integration, and Eq.(\ref{Kinetic equation
for f_0 1}) becomes%
\begin{equation}
\oint dt\frac{e}{2m} Re\left[ E_{y}^{\ast }(x)\frac{df_{1}}{dv_{y}}%
\right] =\oint dtSt(f_{0}).  \label{Kinetic equation for f_0 averaged}
\end{equation}%
Second, we integrate Eq.(\ref{Kinetic equation for f_0 averaged}) over all
possible perpendicular velocities $dv_{y}dv_{z}$ with a given total energy $%
m(v_{y}^{2}+v_{z}^{2})<2\varepsilon $ \cite{Tsendin 77 dc}, \cite{Me and
Tsendin 1992 2}.

Total averaging is a triple integral%
\begin{equation}
\overline{Term(x,\mathbf{v})}\equiv \frac{1}{4\pi }\int \int
dv_{y}dv_{z}\oint dtTerm(x,\mathbf{v}),  \label{averaging}
\end{equation}%
where the factor $1/4\pi $ is introduced for the normalization purposes.
Note, that integral Eq.(\ref{averaging}) is simply averaging over all phase
space available for the electron with the total energy $\varepsilon .$%
\begin{equation}
\frac{1}{4\pi }\int \int dv_{y}dv_{z}\oint dtTerm(x,\mathbf{v})=\frac{m}{%
4\pi }\int dxd^{3}\mathbf{v}\delta \left[ \varepsilon -w-\varphi (x)\right] .
\label{averaging 1}
\end{equation}%
If the $Term(x,\mathbf{v})$ depends on velocity only only via speed $v$,
like ,e.g., the inelastic collision integral Eq.(\ref{inelastic collision
integral}), then integration of Eq.(\ref{averaging 1}) in spherical
coordinates gives%
\begin{eqnarray}
\overline{Term(x,\mathbf{v})}(\varepsilon )
&=&\int_{x_{-}}^{x_{+}}dxv(x,\varepsilon )Term[x,v(x,\varepsilon )],
\label{averaging v} \\
v(x,\varepsilon ) &=&\sqrt{2[\varepsilon -\varphi (x)]/m}.
\end{eqnarray}%
Thus, averaging of the terms, which are functions of the position and the
kinetic energy only, reduces to the integrating over the entire available
discharge volume weighted with the velocity, which is the standard procedure
that also appears in the averaged kinetic equations for the local
(collisional) case \cite{Uwe review}.

\subsection{Calculation of the nonlocal energy diffusion coefficient}

The averaged energy diffusion term originates from the averaged left hand
side of Eq.(\ref{Kinetic equation for f_0 averaged}), which gives%
\begin{equation}
\overline{\frac{eE_{y}(x,t)}{m}\frac{df_{1}}{dv_{y}}}=\frac{e}{8\pi }\int
dxd^{3}\mathbf{v}\delta \left[ \varepsilon -w-\varphi (x)\right] Re%
\left[ E_{y}^{\ast }(x)\frac{df_{1}}{dv_{y}}\right] .
\label{averaging heating 1}
\end{equation}%
Using chain rule for the integration in $dv_{y}$ and the fact that $d\delta
(\varepsilon -w-\varphi )/dv_{y}=mv_{y}d\delta (\varepsilon -w-\varphi
)/d\varepsilon $, Eq.(\ref{averaging heating 1}) becomes

\begin{equation}
\overline{\frac{eE_{y}(x,t)}{m}\frac{df_{1}}{dv_{y}}}=\frac{em}{8\pi }\frac{d%
}{d\varepsilon } Re\int dxd^{3}\mathbf{v}\delta \left[ \varepsilon
-w-\varphi (x)\right] v_{y}E_{y}^{\ast }(x)f_{1}.
\end{equation}%
Substituting $f_{1}$ from Eq.(\ref{f1s paper}) and integrating in the
velocities $v_{y}$ and $v_{z}$ yields
\begin{equation}
\overline{\frac{eE_{y}(x,t)}{m}\frac{df_{1}}{dv_{y}}}=-\frac{d}{d\varepsilon
}D_{\varepsilon }\frac{df_{0}}{d\varepsilon },
\end{equation}%
where $D_{\varepsilon }$ is the energy diffusion coefficient
\begin{equation}
D_{\varepsilon }=\frac{e}{4m} Re\int_{0}^{\varepsilon
}d\varepsilon _{x}\left( \varepsilon -\varepsilon _{x}\right)
\int_{x_{-}(\varepsilon _{x})}^{x_{+}(\varepsilon
_{x})}\frac{dx}{v_{x}}E_{y}^{\ast }(x)V_{y}^{rf}(x,\varepsilon
_{x}). \label{energy diffusion coefficient general}
\end{equation}%
Equation (\ref{energy diffusion coefficient general}) is the general
expression for the energy diffusion coefficient: in the limiting regime of
the small mean free path ($\lambda <<\delta $) it tends to the local limit,
in the intermediate pressure range ($\delta <<\lambda <<L$) Eq.(\ref{energy
diffusion coefficient general}) corresponds to the hybrid heating: electron
motion in the skin layer is collisionless, but the randomization of the
velocity kick acquired during a single pass through the skin layer occurs
due to collisions in the plasma bulk, and in the opposite limit of the large
mean free path ($\lambda >L$) Eq.(\ref{energy diffusion coefficient general}%
) describes collisionless heating (see Appendix B,C for details). If the
collision frequency does not depend on the kinetic energy the direct
substitution of Eq.(\ref{Vrf(x,e)}) for $V_{y}^{rf}$ gives
\begin{equation}
D_{\varepsilon }(\varepsilon )=\frac{\pi }{4}\sum_{n=-\infty }^{\infty
}\int_{0}^{\varepsilon }d\varepsilon _{x}\left\vert E_{yn}(\varepsilon
_{x})\right\vert ^{2}\frac{\varepsilon -\varepsilon _{x}}{\Omega
_{b}(\varepsilon _{x})}\frac{\nu }{\left[ \Omega _{b}(\varepsilon
_{x})n-\omega \right] ^{2}+\nu ^{2}},
\label{energy diffusion coefficient nu constant}
\end{equation}%
where%
\begin{equation}
E_{yn}(\varepsilon _{x})=\frac{1}{\pi }\left[ \int_{0}^{\pi }E_{y}(\theta
)\cos \left( n\theta \right) d\theta \right] .  \label{En paper}
\end{equation}%
Note that $D_{\varepsilon }(\varepsilon )$ in the last equation accounts for
the bounce resonance $\Omega _{b}(\varepsilon _{x})n=\omega $ and the
transit time resonance $\omega =v/\delta $, which corresponds to the maxima
of $E_{yn}(\varepsilon _{x})$.

\section{Self-consistent system of equations}

In summary, the self-consistent system of equations for the kinetic
description of low-pressure discharges accounting for nonlocal and
collisionless electron dynamics contains:

1. The averaged kinetic equation for $f_{0}$ reads%
\begin{equation}
-\frac{d}{d\varepsilon }\left( D_{\varepsilon }+\overline{D_{ee}}\right)
\frac{df_{0}}{d\varepsilon }-\frac{d}{d\varepsilon }\left[ \overline{V_{ee}}+%
\overline{V_{el}}\right] f_{0}=\sum_{k}\left[ \overline{\nu _{k}^{\ast }%
\frac{\sqrt{(w+\varepsilon _{k}^{\ast })}}{\sqrt{w}}}f_{0}(\varepsilon
+\varepsilon _{k}^{\ast })-\overline{\nu _{k}^{\ast }}f_{0}\right] ,
\label{final averaged k.e.}
\end{equation}%
where the upper bar denotes averaging according to Eq.(\ref{averaging v})
and $D_{ee}$ is given by Eq.(\ref{Dee}), $V_{ee}$ by Eq.(\ref{Vee}) $V_{el}$
by Eq.(\ref{vel}), and $D_{\varepsilon }$ by Eq.(\ref{energy diffusion
coefficient general}) or by Eq.(\ref{energy diffusion coefficient nu
constant}).

2. The rf electric field is determined from the Maxwell Eq.(\ref{Maxwell eqs}%
), where the electron current is given by Eq.(\ref{J=G(xx')}).

3. The electrostatic potential is obtained using the quasineutrality
condition
\begin{equation}
n_{i}(x)=\int_{\varphi (x)}^{\infty }f_{0}(\varepsilon )\sqrt{\varepsilon
-\varphi (x)}d\varepsilon ,  \label{quasineutrality}
\end{equation}%
where $n_{i}(x)$ is the ion density profile given by a set of number
particles and ion momentum equations \cite{badri and me}. Eq.(\ref%
{quasineutrality}) is solved in the form of a differential equation \cite{My
CCP}
\begin{equation}
\frac{d\varphi }{dx}=-T_{e}^{scr}(x)\frac{d\ln [n_{i}(x)]}{dx},
\end{equation}%
where $T_{e}^{scr}(x)$ is the electron screening temperature%
\begin{equation}
T_{e}^{scr}(x)=\left[ \frac{1}{2n(x)}\int_{\varphi (x)}^{\infty
}f_{0}(\varepsilon )\frac{d\varepsilon }{\sqrt{\varepsilon -\varphi (x)}}%
\right] ^{-1}.
\end{equation}

4. The power deposition can be computed as
\begin{equation}
P(x)=\frac{1}{2} Re\left[ E_{y}^{\ast }(x)j(x)\right] .
\label{P(j)}
\end{equation}

Substituting Eq.(\ref{j(G)2}) and changing the integration order, Eq.(\ref%
{P(j)}) becomes%
\begin{equation}
P=-\sqrt{2m}\int_{0}^{\infty }D_{\varepsilon }(\varepsilon )\frac{%
df_{0}(\varepsilon )}{d\varepsilon }d\varepsilon .  \label{P(D)}
\end{equation}%
Equation (\ref{P(D)}) can be used as a consistency check.

This research was supported by the U.S. Department of Energy Office of
Fusion Energy Sciences through a University Research Support Program. The
author gratefully acknowledge helpful discussions with R. Davidson, Badri
Ramamurthi and E. Startsev.

\section{Appendixes}

\subsection{Derivation of $f_{1}$}

Direct integration of Eq.(\ref{Vlasov 1}) yields%
\begin{equation}
f_{1+}(x,v)=-ev_{y}\frac{df_{0}}{d\varepsilon }\left[ \int_{x_{-}}^{x}e^{-[%
\Phi (x)-\Phi (x^{\prime })]}E^{\prime }d\tau ^{\prime
}+f_{1+}(x_{-})e^{-\Phi (x)}\right] .  \label{f1+}
\end{equation}%
\begin{equation}
f_{1-}(x,v)=-ev_{y}\frac{df_{0}}{d\varepsilon }\left[ -\int_{x_{-}}^{x}e^{%
\Phi (x)-\Phi (x^{\prime })}E_{y}{}^{\prime }d\tau ^{\prime
}+f_{1-}(x_{-})e^{\Phi (x)}\right] .  \label{f1-}
\end{equation}%
where $\pm $ signs denote $v_{x}>0$ and $v_{x}<0$, respectively, and for
brevity we introduced $E_{y}{}^{\prime }\equiv E_{y}(x^{\prime })$ and $%
d\tau =dx/|v_{x}|$. The two constants $f_{1+}(x_{-}),f_{1-}(x_{-})$ are to
be determined from the boundary condition at the turning points, and%
\begin{equation}
\Phi (x)\equiv \int_{x_{-}}^{x}(-i\omega +\nu )d\tau .  \label{Phase}
\end{equation}%
Due to continuity of the EVDF
\begin{equation}
f_{1-}(x_{-})=f_{1+}(x_{-}),\quad f_{1-}(x_{+})=f_{1+}(x_{-}).  \label{b.c.}
\end{equation}%
Substituting the boundary condition at the turning points Eqs.(\ref{b.c.})
into Eqs.(\ref{f1+}) and (\ref{f1-}) yields

\[
-\int_{x_{-}}^{x_{+}}e^{\Phi -\Phi ^{\prime }}E_{y}{}^{\prime }d\tau
^{\prime }+f_{1-}(x_{-})e^{\Phi _{+}}=\int_{x_{-}}^{x_{+}}e^{-(\Phi
_{+}-\Phi ^{\prime })}E_{y}{}^{\prime }d\tau ^{\prime
}+f_{1+}(x_{-})e^{-\Phi _{+}},
\]%
or
\begin{equation}
f_{1+}(x_{-})=\frac{e}{\sinh \Phi _{+}}\int_{x_{-}}^{x_{+}}\cosh (\Phi
_{+}-\Phi ^{\prime })E_{y}{}^{\prime }d\tau ^{\prime },  \label{f1(x-)}
\end{equation}%
where $\Phi \equiv \Phi (x)$, $\Phi ^{\prime }\equiv \Phi (x^{\prime })$,
and $\Phi _{+}\equiv \Phi (x_{+})$. $f_{1}$ enters into the current
calculation only as a sum $f_{1+}+f_{1-}$. Therefore, we compute $%
f_{1s}\equiv 1/2(f_{1+}+f_{1-})$ from Eqs.(\ref{f1+}) and (\ref{f1-})
\begin{equation}
f_{1s}=-ev_{y}\frac{df_{0}}{d\varepsilon }\left\{ f_{1+}(x_{-})\cosh \Phi
-\int_{x_{-}}^{x}\sinh (\Phi -\Phi ^{\prime })E_{y}(\theta ^{\prime })d\tau
^{\prime }\right\} ,  \label{f1s 1}
\end{equation}%
substituting $f_{1+}(x_{-})$ from Eq.(\ref{f1(x-)}) gives%
\begin{equation}
f_{1s}=-mv_{y}V_{y}^{rf}\frac{df_{0}}{d\varepsilon },
\end{equation}%
where
\begin{equation}
V_{y}^{rf}=\frac{e}{m}\frac{1}{\sinh \Phi _{+}}\left[
\begin{array}{c}
\cosh \Phi \int_{x_{-}}^{x_{+}}\cosh (\Phi _{+}-\Phi ^{\prime
})E_{y}{}^{\prime }d\tau ^{\prime }- \\
\sinh \Phi _{+}\int_{x_{-}}^{x}\sinh (\Phi -\Phi ^{\prime })E_{y}{}^{\prime
}d\tau ^{\prime }%
\end{array}%
\right] .  \label{f1s 2}
\end{equation}%
Splitting the first term into two integrals $\int_{x_{-}}^{x_{+}}=%
\int_{x_{-}}^{x}+\int_{x}^{x_{+}},$ and accounting for the fact that%
\begin{equation}
\cosh \Phi \cosh (\Phi _{+}-\Phi ^{\prime })-\sinh \Phi _{+}\sinh (\Phi
-\Phi ^{\prime })=\cosh \Phi ^{\prime }\cosh (\Phi _{+}-\Phi ^{\prime })
\end{equation}%
gives%
\begin{equation}
V_{y}^{rf}=\frac{e}{m}\frac{1}{\sinh \Phi _{+}}\left[
\begin{array}{c}
\cosh \Phi \int_{x}^{x_{+}}E_{y}{}^{\prime }\cosh (\Phi _{+}-\Phi ^{\prime
})d\tau ^{\prime }+ \\
\cosh (\Phi _{+}-\Phi )\int_{x_{-}}^{x}E_{y}{}^{\prime }\cosh \Phi ^{\prime
}d\tau ^{\prime }%
\end{array}%
\right] .  \label{f1s final}
\end{equation}

\subsection{Diffusion coefficient in the energy space}

The equation for the energy diffusion coefficient%
\begin{equation}
D_{\varepsilon }=\frac{e}{4m} Re\int_{0}^{\varepsilon
}d\varepsilon
_{x}\left( \varepsilon -\varepsilon _{x}\right) \int_{x_{-}}^{x_{+}}\frac{dx%
}{v_{x}}E_{y}^{\ast }(x)V_{y}^{rf}(x,\varepsilon _{x})
\label{energy diffusion coefficient}
\end{equation}%
has correct limits in local and nonlocal cases.

\subsubsection{local limit $\protect\lambda <<L$}

In the local limit of the small mean free path, substituting $V_{y}^{rf}$
from Eq.(\ref{Vrf local}) into Eq.(\ref{energy diffusion coefficient}) gives

\begin{equation}
D_{\varepsilon }=\frac{e^{2}}{4m^{2}} Re\int_{0}^{\varepsilon
}d\varepsilon _{x}\left( \varepsilon -\varepsilon _{x}\right)
\int_{x_{-}}^{x_{+}}\frac{dx}{\sqrt{2(\varepsilon _{x}-\varphi )/m}}\frac{%
E_{y}^{\ast }(x)E_{y}(x)}{-i\omega +\nu }.
\label{energy diffusion coefficient local}
\end{equation}%
Changing the order of the integration and accounting for the fact that
\[
\frac{1}{m^{2}}\int_{0}^{\varepsilon }d\varepsilon _{x}\frac{\varepsilon
-\varepsilon _{x}}{\sqrt{2(\varepsilon _{x}-\varphi )/m}}=\frac{2}{3}v^{3}
\]%
\begin{equation}
D_{\varepsilon }=\frac{e^{2}}{6} Re\int_{x_{-}}^{x_{+}}dx|E_{y}|^{2}%
\frac{\nu v^{3}}{(\omega ^{2}+\nu ^{2})},
\label{energy diffusion coefficient local final}
\end{equation}%
which corresponds to the local limit \cite{Kolobov and Godyak review},\cite%
{Uwe review}.

\subsubsection{nonlocal limit $\protect\delta <<\protect\lambda <<L$}

In the nonlocal limit, collisions during the electron motion in the skin
layer are rare. Therefore, $V_{y\pm }^{rf}$ is simply velocity "kick" due
the rf electric field. Recalling that $V_{y}^{rf}(x,\varepsilon
)=(V_{y+}^{rf}+V_{y-}^{rf})/2$, the last factor in Eq.(\ref{energy diffusion
coefficient}) can be written as
\begin{equation}
\frac{1}{2}\frac{e}{m}\int_{x_{-}}^{x_{+}}\frac{dx}{v_{x}}E_{y}^{\ast
}(x)V_{y}^{rf}(x,\varepsilon _{x})=\left\langle \frac{1}{2}\oint
d\tau \frac{d\Delta V_{y}(\tau )}{d\tau }\Delta V_{y}(\tau )\right\rangle =\frac{1%
}{4}\left\langle \Delta V_{y\infty }^{2}\right\rangle ,
\end{equation}%
where $\oint d\tau $ is an integral along the electron trajectory
entering and leaving the skin layer, and $\Delta V_{y\infty }$ is
the total velocity kick after a single path through the skin
layer, and the angular brackets denote averaging over phases of
the rf field. Eq.(\ref{energy diffusion coefficient}) simplifies
to
\begin{equation}
D_{\varepsilon }=\frac{1}{8}\int_{0}^{\varepsilon }d\varepsilon _{x}\left(
\varepsilon -\varepsilon _{x}\right) \left\langle \Delta V_{y\infty
}^{2}\right\rangle .  \label{D nonlocal}
\end{equation}%
In the limit of a uniform plasma Eq.(\ref{D nonlocal}) was proposed in Ref.
\cite{me APL}, \cite{Kolobov PRE 55}.

\subsubsection{collisionless limit $\protect\lambda >>L$}

The energy diffusion coefficient Eq.(\ref{energy diffusion coefficient})
with substitution of $V_{y}^{rf}$ from Eq.(\ref{f1s final}) is determined by
the following integral:

$Int=\int_{x_{-}}^{x_{+}}E_{y}^{\ast }(x)V_{y}^{rf}d\tau =\frac{1}{\sinh
\Phi _{+}}\int_{x_{-}}^{x_{+}}E_{y}^{\ast }(x)d\tau $

$\left[ \cosh \Phi \int_{x}^{x_{+}}E_{y}(x^{\prime })\cosh (\Phi _{+}-\Phi
^{\prime })d\tau ^{\prime }+\cosh (\Phi _{+}-\Phi
)\int_{x_{-}}^{x}E_{y}(\theta ^{\prime })\cosh \Phi ^{\prime }d\tau ^{\prime
}\right] .$

The term in brackets is $\cosh \Phi _{+}\cosh \Phi
\int_{x_{-}}^{x_{+}}E_{y}\cosh \Phi ^{\prime }d\tau ^{\prime }+$

$\sinh \Phi _{+}\left[ \cosh \Phi \int_{x}^{x_{+}}E_{y}\sinh \Phi ^{\prime
}d\tau ^{\prime }+\sinh \Phi \int_{x_{-}}^{x}E_{y}\cosh \Phi ^{\prime }d\tau
^{\prime }\right] $.

Therefore, $Int=\frac{\cosh \Phi _{+}}{\sinh \Phi _{+}}%
\int_{x_{-}}^{x_{+}}E_{y}^{\prime }\cosh \Phi ^{\prime }d\tau ^{\prime
}\int_{x_{-}}^{x_{+}}E_{y}^{\ast \prime }\cosh \Phi ^{\prime }d\tau ^{\prime
}+Int1$, where

$Int1=\int_{x_{-}}^{x_{+}}E_{y}^{\ast }d\tau \left[ \cosh \Phi
\int_{x}^{x_{+}}E_{y}^{\prime }\sinh \Phi ^{\prime }d\tau ^{\prime }+\sinh
\Phi \int_{x_{-}}^{x}E_{y}^{\prime }\cosh \Phi ^{\prime }d\tau ^{\prime }%
\right] $ . Integrating in parts gives $Int1=\int_{x_{-}}^{x_{+}}\sinh \Phi
d\tau \left[ \int_{x_{-}}^{x}\left[ E_{y}{}^{\prime }E_{y}^{\ast
}+E_{y}E_{y}^{\ast }\right] \cosh \Phi ^{\prime }d\tau ^{\prime }\right] $

In the nonlocal $\sinh \Phi =i\sin \omega \tau +\nu \tau \cos \omega \tau .$
Because the energy diffusion coefficient is determined by the real part of
the integral and the real part of the phase is small ($\sim \nu $ ) $Int1$
can be neglected. Therefore,
\begin{equation}
D_{\varepsilon }=\frac{1}{4}\ Re\int_{0}^{\varepsilon
}d\varepsilon _{x}\left( \varepsilon -\varepsilon _{x}\right)
\coth \Phi _{+}\int_{x_{-}}^{x_{+}}E_{y}{}^{\prime }\cos \omega
\tau ^{\prime }d\tau ^{\prime }\int_{x_{-}}^{x_{+}}E_{y}^{\ast
}{}^{\prime \prime }\cos \omega \tau ^{\prime \prime }d\tau
^{\prime \prime },
\end{equation}%
where $\sinh \Phi _{+}=i\sin \omega T+\nu T\cos \omega T$. Main contribution
comes from the points, where $\omega T=\pi n$ and $\coth \Phi _{+}=\pi
\sum_{n=1}\delta (\omega T-\pi n)$.

\begin{eqnarray}
D_{\varepsilon }(\varepsilon ) &=&\frac{\pi e^{2}}{4}\sum_{n=-\infty
}^{\infty }\int_{0}^{\varepsilon }d\varepsilon _{x}\left\vert Ef\right\vert
^{2}\left( \varepsilon -\varepsilon _{x}\right) \delta \left[ \omega
T(\varepsilon _{x})-\pi n\right] ,  \label{D Aliev} \\
Ef(\varepsilon _{x}) &=&\int_{x_{-}(\varepsilon _{x})}^{x_{+(\varepsilon
_{x})}}E_{y}(x^{\prime })\cos \omega \tau ^{\prime }d\tau ^{\prime }
\end{eqnarray}%
This corresponds to the pervious results \cite{Aliev and me}.

\subsection{Alternative derivations in Fourier space.}

The direct calculation described in the previous sections are rather
cumbersome. The alternative derivation can be done easier using Fourier
series.

It is convenient to introduce the variable angle of the bounce motion%
\begin{equation}
\theta (x,\varepsilon _{x})=\frac{\pi sgn(v_{x})}{T(\varepsilon _{x})}%
\int_{x_{-}}^{x}\frac{dx}{\left\vert v_{x}(\varepsilon _{x})\right\vert },
\label{angle}
\end{equation}%
where $T$ is the half of the bounce period for the electron motion in the
potential well $\varphi (x)$, which is given by
\begin{equation}
T(\varepsilon _{x})=\int_{x_{-}}^{x_{+}}\frac{dx}{\left\vert
v_{x}(\varepsilon _{x})\right\vert }.  \label{bounce period}
\end{equation}%
The bounce frequency for the electron in the potential well is $\Omega
_{b}(\varepsilon _{x})=\pi /T(\varepsilon _{x})$. Utilizing angle variable,
Eq.(\ref{Vlasov 1}) simplifies to become%
\begin{equation}
-i\omega f_{1}+\Omega _{b}\frac{\partial f_{1}}{\partial \theta }%
|_{\varepsilon _{x}}+v_{y}eE_{y}(\theta )\frac{df_{0}}{d\varepsilon }=-\nu
f_{1},  \label{Vlasov 2}
\end{equation}%
where $\pm $ signs denote $v_{x}>0$ and $v_{x}<0$, respectively and $\Omega
_{b}(\varepsilon _{x})=\pi /T(\varepsilon _{x})$ is the bounce frequency in
the potential well.

We shall use Fourier series in variable $\theta $:%
\begin{equation}
g(x,\varepsilon _{x})=\sum_{n=-\infty }^{\infty }g_{n}\exp \left( in\theta
\right)  \label{Fourier direct}
\end{equation}%
\begin{equation}
g_{n}=\frac{1}{2\pi }\left[ \int_{-\pi }^{\pi }g(\theta ,\varepsilon
_{x})\exp \left( -in\pi \theta \right) d\theta \right] .
\label{Fourier back}
\end{equation}%
Note that in the last integral, the region $0<\theta <\pi $ corresponds to $%
v_{x}>0$ , and the region $-\pi <\theta <0$ corresponds to $v_{x}<0$.
Utilizing the Fourier series Eq.(\ref{Fourier back}), the Vlasov equation
becomes%
\begin{equation}
(in\Omega _{b}-i\omega +\nu )f_{1n}=-E_{yn}v_{y}\frac{df_{0}}{d\varepsilon },
\label{Vlasov Fourier}
\end{equation}%
where
\begin{equation}
E_{yn}(\varepsilon _{x})=\frac{1}{\pi }\left[ \int_{0}^{\pi }E_{y}(\theta
)\cos \left( n\theta \right) d\theta \right] .  \label{En}
\end{equation}%
Making use of Fourier series Eq.(\ref{Fourier direct}), Eq.(\ref{Vlasov
Fourier}) gives%
\[
f_{1s}(x,\varepsilon _{x})=-mv_{y}V_{y}^{rf}(x,\varepsilon _{x})\frac{df_{0}%
}{d\varepsilon },
\]%
\begin{equation}
V_{y}^{rf}(x,\varepsilon _{x})=\frac{e}{m}\sum_{n=-\infty }^{\infty }\frac{%
E_{yn}\cos [n\theta (x)]}{in\Omega _{b}-i\omega +\nu }.  \label{f1s Fourier}
\end{equation}%
Eq.(\ref{f1s Fourier}) is the alternative form of Eq.(\ref{f1s final}).

Substituting the function $V_{y}^{rf}(x,\varepsilon _{x})$ from Eq.(\ref{f1s
Fourier}) into Eq.(\ref{j(A)final}) gives the current density
\begin{equation}
j(x)=\frac{e^{2}}{2m}\sum_{n=-\infty }^{\infty }\int_{\varphi (x)}^{\infty }%
\frac{\Gamma (\varepsilon )}{\sqrt{\varepsilon -\varphi (x)}}\frac{%
E_{yn}\cos [n\theta (x)]}{in\Omega _{b}-i\omega +\nu }d\varepsilon .
\label{j as Fourier series}
\end{equation}%
The averaged energy coefficient is given by Eq.(\ref{energy diffusion
coefficient}). Substituting the function $V_{y}^{rf}(x,\varepsilon _{x})$
from Eq.(\ref{f1s Fourier}) into Eq.(\ref{energy diffusion coefficient})
gives

\[
D_{\varepsilon }=\frac{e^{2}}{4} Re\int_{0}^{\varepsilon
}d\varepsilon
_{x}\left( \varepsilon -\varepsilon _{x}\right) \int_{x_{-}}^{x_{+}}\frac{dx%
}{v_{x}}E_{y}^{\ast }(x)\sum_{n=-\infty }^{\infty }\frac{E_{yn}\cos [n\theta
(x)]}{in\Omega _{b}-i\omega +\nu },
\]%
or%
\begin{equation}
D_{\varepsilon }(\varepsilon )=\frac{\pi e^{2}}{4}\sum_{n=-\infty }^{\infty
}\int_{0}^{\varepsilon }d\varepsilon _{x}\left\vert E_{yn}(\varepsilon
_{x})\right\vert ^{2}\left( \varepsilon -\varepsilon _{x}\right) \frac{\nu }{%
\Omega _{b}(\varepsilon _{x})\left[ \Omega _{b}(\varepsilon _{x})n-\omega %
\right] ^{2}+\nu ^{2}}.  \label{Energy diffusion Fourier final}
\end{equation}%
Note that Eq.(\ref{Energy diffusion Fourier final}) is valid for any
collision frequency, and Eq.(\ref{D Aliev}) is valid only for $\nu <<\omega $%
.

\subsection{Solving the Maxwell equations for the rf electric field using
Fourier Series}

System that has an antenna at $x=0$ and a grounded electrode at $x=L$ in the
uniform plasma were studied in Ref.\cite{Shaing}. The papers \cite{Blevin
theory} and \cite{Yoon} considered a cylindrical-like system in the uniform
plasma. Both papers used Fourier series to solve Maxwell's equations. Here,
we shall generalize the procedure for a case of a nonuniform plasma.

Similarly to the previous analysis, it is convenient to continue the rf
electric field symmetrically $E_{y}(x)=E_{y}(-x)$ outside of the slab. Then,
the electric field is given by Fourier series \cite{Shaing}:

\begin{equation}
E_{y}(x)=\sum_{s=0}^{\infty }\Xi _{s}\cos (k_{s}x),  \label{Fourier series E}
\end{equation}%
where $s$ is an integer, $k_{s}=(2s+1)\pi /(2L),$ for the case of the
grounded electrode, and $k_{s}=(2s+1)\pi /L,$ for the case of the
cylindrical-like system. Substituting (\ref{Fourier series E}) into Eq.(\ref%
{Maxwell eqs}) and integrating with the weight $\cos (k_{s}x)/L$ over the
region $[-L,L]$ yields%
\begin{equation}
\left( -k_{s}^{2}+\frac{\omega ^{2}}{c^{2}}\right) \Xi _{s}=-\frac{4\pi
i\omega }{c^{2}}\left[ j_{s}+\frac{I\left[ 1+\delta _{anti,k}\right] }{L}%
\right] ,  \label{Eq. for Furier E}
\end{equation}%
where
\begin{equation}
j_{s}=\frac{2}{L}\int_{0}^{L}j(x)\cos (k_{s}x)dx.  \label{Fourier series J}
\end{equation}%
Substituting the equation for the current density Eq.(\ref{j as Fourier
series}) gives
\begin{equation}
j_{s}=\frac{ne^{2}}{m}\frac{1}{s\Omega _{bT}}\sum_{l=0}^{\infty }\Xi
_{l}Z_{s,l}^{gen}\left( \frac{\omega +i\nu }{s\Omega _{bT}}\right) ,
\label{J as double summ}
\end{equation}%
where $\Omega _{bT}=v_{T}\pi /L$, and we introduced the generalized plasma
dielectric function
\begin{equation}
Z_{s,l}^{gen}\left[ \xi =\frac{\omega +i\nu }{s\Omega _{bT}}\right] \equiv
\sqrt{\frac{2}{m}}\frac{s\Omega _{bT}}{L}\sum_{n=-\infty }^{\infty
}\int_{0}^{\infty }\frac{\Gamma (\varepsilon )}{in\Omega _{b}(\varepsilon
_{x})-is\Omega _{bT}\xi }\frac{\pi }{\Omega _{b}(\varepsilon _{x})}%
G_{s,n}(\varepsilon )G_{l,n}(\varepsilon )d\varepsilon ,
\label{plasma dielectric function}
\end{equation}%
where the coefficients $G_{l,n}(\varepsilon )$ are the Fourier transform of $%
\cos (k_{l}x)$ in the bounce motion of the electron in potential well:%
\begin{equation}
G_{l,n}(\varepsilon )=\frac{1}{T}\left[ \int_{0}^{T}\cos [k_{l}x(\tau )]\cos
\left( \frac{\pi n\tau }{T}\right) d\tau \right] .  \label{G coefficients}
\end{equation}%
In the limit of a uniform plasma $\tau =x/v_{x}$, $\tau =L/v_{x}$ which gives%
\begin{equation}
G_{l,n}(\varepsilon )=\frac{1}{L}\left[ \int_{0}^{L}\cos (k_{l}x)\cos \left(
\frac{n\pi x}{L}\right) dx\right] .  \label{G cfnts uniform pl.}
\end{equation}%
For a cylindrical-like system coefficients $G_{l,n}(\varepsilon )$ are
particular simple
\begin{equation}
G_{l,n}(\varepsilon )=\frac{1}{2}\delta _{|l|,|n|},
\end{equation}%
and the generalized plasma dielectric function is
\begin{equation}
Z_{s,l}^{gen}(\xi )=\delta _{s,l}Z(\xi ),
\end{equation}%
where $Z(\xi )$ is the "standard" plasma dielectric function
\begin{equation}
Z(\xi )=\pi ^{-1/2}\int_{-\infty }^{\infty }dt\frac{\exp (-t^{2})}{t-\xi }.
\end{equation}%
Eq.(\ref{J as double summ}) is identical to the results of Ref.\cite{Blevin
theory} for cylindrical-like configuration uniform plasma with a Maxwellian
EVDF.

Coefficients $G_{l,n}(\varepsilon )$ can be effectively computed using the
fast Fourier transform \cite{Startsev}. The off-diagonal coefficients are
generally very small, that is why using this spectral method makes computing
much faster than the straight forward finite difference method used in Ref.%
\cite{badri and me}.

\end{document}